\documentclass{ws-ijmpe}
\usepackage[super,compress]{cite}
\usepackage{url}

\newcommand{\be}{\begin{equation}}
\newcommand{\ee}{\end{equation}}
\newcommand{\beq}{\begin{eqnarray}}
\newcommand{\eeq}{\end{eqnarray}}
\begin{document}

\markboth{Brynmor Haskell}{R-modes in neutron stars}

\catchline{}{}{}{}{}

\title{R-modes in neutron stars: theory and observations}

\author{B. HASKELL}

\address{School of Physics, The University of Melbourne,\\
Parkville, VIC 3010,
Australia\\
brynmor.haskell@unimelb.edu.au}



\maketitle

\begin{history}
\received{Day Month Year}
\revised{Day Month Year}
\end{history}

\begin{abstract}

In this article I will review the theory behind the gravitational wave driven r-mode instability in rapidly rotating neutron stars and discuss which constraints can be derived from observations of spins and temperatures in Low Mass X-ray Binaries. I will discuss how a standard, `minimal' neutron star model is not consistent with the data, and discuss some of the additional physical mechanisms that could reconcile theory with observations. In particular I will focus on additional forms of damping due to exotic cores and on strong mutual friction due to superfluid vortices cutting through superconducting flux tubes, and examine the repercussions these effects could have on the saturation amplitude of the mode. Finally I will also discuss the possibility that oscillations due to r-modes may have been recently observed in the X-ray light curves of two Low Mass X-ray Binaries.

\end{abstract}


\ccode{PACS numbers:95.85.Sz, 97.60.Jd, 97.10.Sj}


\section{Introduction}

Rapidly rotating Neutron Stars (NSs) are one of the main targets for Gravitational Wave (GW) detectors such as Advanced LIGO and Virgo \cite{Riles}, as there are several mechanisms that could lead to non-axisymmetric deformations of the star and ultimately to a continuous GW signal \cite{book}. In particular there are several modes of oscillation of the NS that can result in GW emission, and in this context the so-called r-mode has attracted considerable attention, as it is generically unstable to GW emission \cite{Nils98,Morsink}, and could thus grow to large amplitudes and offer the best detection prospects.

The possibility of detecting GWs from NS modes of oscillation is particularly exciting as gravitational wave asteroseismology would allow us to probe the interior structure and composition of NSs in great detail, in the same way as electromagnetic asteroseismology has, in recent years, allowed us to significantly enhance our understanding of many other stars, from white dwarfs to red giants \cite{WDrev, Astro1}.

Let us remind the reader that, with a mass roughly equal to that of the sun compressed in a 10 km radius, NSs are one of the most compact objects in the universe and their interior density can easily surpass nuclear saturation density. Furthermore NSs are cold objects as, although their core temperatures are in the region of $T\approx 10^8$ K, the thermal energy is small compared to the Fermi energy in most of the star. This can have important consequences for the dynamics of the system, as  neutrons will pair and form a superfluid, while protons will be superconducting.
Probing the interior structure of these objects would allow us to probe physics in an entirely different regime from ground based experiments, such as heavy ion colliders, which generally probe the low density and high temperature section of the QCD phase diagram \cite{Brambilla}. While at asymptotically high densities quarks are though to pair in the so-called Colour-Flavour-Locked (CFL) phase \cite{AlfordCFL}, for realistic NS densities the ground state of matter is unknown, and only astrophysical observations will be able to solve this problem.

In order to obtain astrophysical constraints and set the theoretical basis for the analysis of future GW data, in this article we will focus on some of the most promising systems for GW detection: Low Mass X-ray Binaries (LMXBs). These are binary systems in which a compact object (in our case a neutron star) is accreting matter from a less evolved companion that fills its Roche lobe. As matter is transferred from the secondary to the primary it forms an accretion disc and eventually is accreted by the NSs, spinning it up. This is thought to be the mechanism by which old, long period, stars are recycled to millisecond periods and millisecond radio pulsars are eventually formed \cite{Alpar82}. The reason LMXBs are invoked as GW sources lies in an observational puzzle: while accretion should be able to spin-up the NS to its Keplerian break-up frequency (which is equation of state dependent, but generally above 1.5 KHz), there appears to be a cutoff at around 700 Hz in the distribution of observed spins for both the LMXBs and the millisecond radio pulsars. It was thus suggested that GW emission could provide an additional spin-down torque that would balance the torque due to accretion and stall the spin-up \cite{PP78, Bild98}. Several mechanisms have been proposed, including `mountains'  supported by the crust \cite{Bild98}, the core \cite{Ben,CFL1} or confined by the magnetic field \cite{Payne05}, and unstable modes of oscillation \cite{Nils98, Morsink}. Although detection will be challenging for many of these scenarios \cite{Watts08, Mountains15} and in several cases it is more likely that the disc/magnetosphere interaction is dominating the torques \cite{HP, ABCPHD}, these systems still allow for the best constraints on the physics of the r-mode instability and the interior dynamics of NSs. Furthermore recent detections of oscillations in the X-ray light curve of two LMXBs may be interpreted as r-modes perturbing the electromagnetic emission, and open the fascinating prospect of combining GW and electromagnetic signals to study NS interiors \cite{Simin1,Simin2}.

\section{The r-mode instability window}

We begin by focusing on the r-mode instability. An r-mode is a fluid mode of oscillation for which the restoring force is the Coriolis force. It thus only exists in a rotating star. In Newtonian gravity, and to first order in the rotational frequency $\Omega$ of the star, it is purely toroidal and for the Eulerian velocity perturbation $\delta\mathbf{v}$ one has (for an in depth review and a discussion of relativistic effects see e.g. references [\refcite{NilsRev}] and [\refcite{NilsGR}]):
\be
\delta\mathbf{v}=\alpha\left(\frac{r}{R}\right)^l R\Omega \mathbf{Y}_{lm}^B e^{i\omega t}
\ee
where $ \mathbf{Y}_{lm}^B=[l(l+1)]^{-1/2}r \nabla\times (r\nabla Y_{lm})$ is the magnetic-type vector spherical harmonic (with $Y_{lm}$ the standard spherical harmonics), $R$ is the stellar radius and $\alpha$ the dimensionless amplitude of the mode \cite{Owen98}, while the frequency of the mode $\omega$ takes the form \cite{NilsRev}:
\be
\omega=\frac{2 m}{l(l+1)}\Omega+O(\Omega^3)
\ee
This mode is interesting for our discussion because it is generically unstable to GW emission. In particular if we examine the pattern speed for an r-mode this is, in a frame rotating with the star:
\be
\sigma_r=-\frac{\omega_r}{m}=-\frac{2\Omega}{l(l+1)}
\ee
In the inertial frame on the other hand one has
\be
\sigma_i=\frac{(l-1)(l+2)}{l(l+1)}\Omega
\ee
so that a mode that is retrograde in the rotating frame appears prograde in the inertial frame. The r-modes thus satisfy the criterion for the so-called Chandrasekhar-Friedman-Schutz (CFL) instability \cite{Chandra70, FS}, which allows for the star to find  lower energy and angular momentum configurations in which the mode amplitude can grow. Note that other modes of oscillation can also be unstable, and another candidate for GW detections is the f-mode\cite{NilsRev}. This mode is not, however, generically unstable, but only goes unstable above a critical frequency. Furthermore in cold systems, such as the LMXBs we consider, the f-mode instability is generally stabilised by viscosity due to superfluid mutual friction \cite{LM95, fmode1,fmode2}.
The strongest contribution to GW emission is due to the $l=m=2$ r-mode an  in this case for an $n=1$ polytrope (which we shall take as our equation of state for all the following estimates) the growth time of the instability is \cite{NilsRev}:
\be
\tau_{gw}=-47 \left(\frac{M}{1.4 M_\odot}\right)^{-1}\left(\frac{R}{10 \mbox{ km}}\right)^{-4}\left(\frac{P}{1 \mbox{ ms}}\right)^6 \mbox{ s},
\ee

Naturally the mode amplitude can only grow provided viscosity cannot damp the instability faster than GW emission can drive it. Viscosity in neutron star interiors is not entirely understood, and it is thus difficult to model in detail. To illustrate the problem we will thus first introduce a `minimal' NS model, corresponding the most commonly considered setup, and calculate the region of parameter space in which the r-mode can grow. In the following we will then show that X-ray observations of spins and temperatures of NSs in LMXBs point towards the need for additional physics in our model.

In our `minimal' model we will assume that, at high temperatures, bulk viscosity due to modified URCA reactions provides the main damping mechanism, while at low temperatures the main contribution is from shear viscosity, due to standard scattering processes (mainly electron-electron in superfluid matter \cite{PhysA}), or from viscosity at the crust-core interface. We assume that there are  no dynamically important magnetic fields or superfluid degrees of freedom, no exotica in the core, and, as previously stated, we take an $n=1$ polytrope as equation of state. With these assumptions we can calculate the damping timescale associated with bulk viscosity, $\tau_{bv}$, and shear viscosity, $\tau_{sv}$ \cite{NilsRev}:
\beq
\tau_{bv}&=&2.7 \times 10^{17} \left(\frac{M}{1.4 M_\odot}\right)\left(\frac{R}{10\mbox{ km}}\right)^{-1}\left(\frac{P}{10^{-3} \mbox{s}}\right)^2\left(\frac{T}{10^8 \mbox{K}}\right)^{-6} s\\
\tau_{sv}&=&2.2 \times 10^5 \left(\frac{M}{1.4 M_\odot}\right)^{-1}\left(\frac{R}{10\mbox{ km}}\right)^5\left(\frac{T}{10^8 \mbox{K}}\right)^2 s
\eeq
where $M$ is the mass of the star, $R$ is the radius, $P$ is the rotation period and $T$ is the core temperature. For damping due to Ekman pumping at the crust/core interface we use the estimate of Glampedakis \& Andersson \cite{Ek1, Ek2} with a slippage parameter $\mathcal{S}=0.05$. The slippage parameter is essentially the ratio between the curst/core velocity difference and the mode velocity, and accounts for the fact that the crust is not completely rigid, but can participate in the oscillation. In this case the damping timescale is:
\be
\tau_{ek}=3\times 10^4 \left(\frac{P}{10^{-3} \mbox{s}}\right)^{1/2}\left(\frac{T}{10^8 \mbox{K}}\right) s
\ee

\begin{figure}[th]
\centerline{\includegraphics[scale=.42]{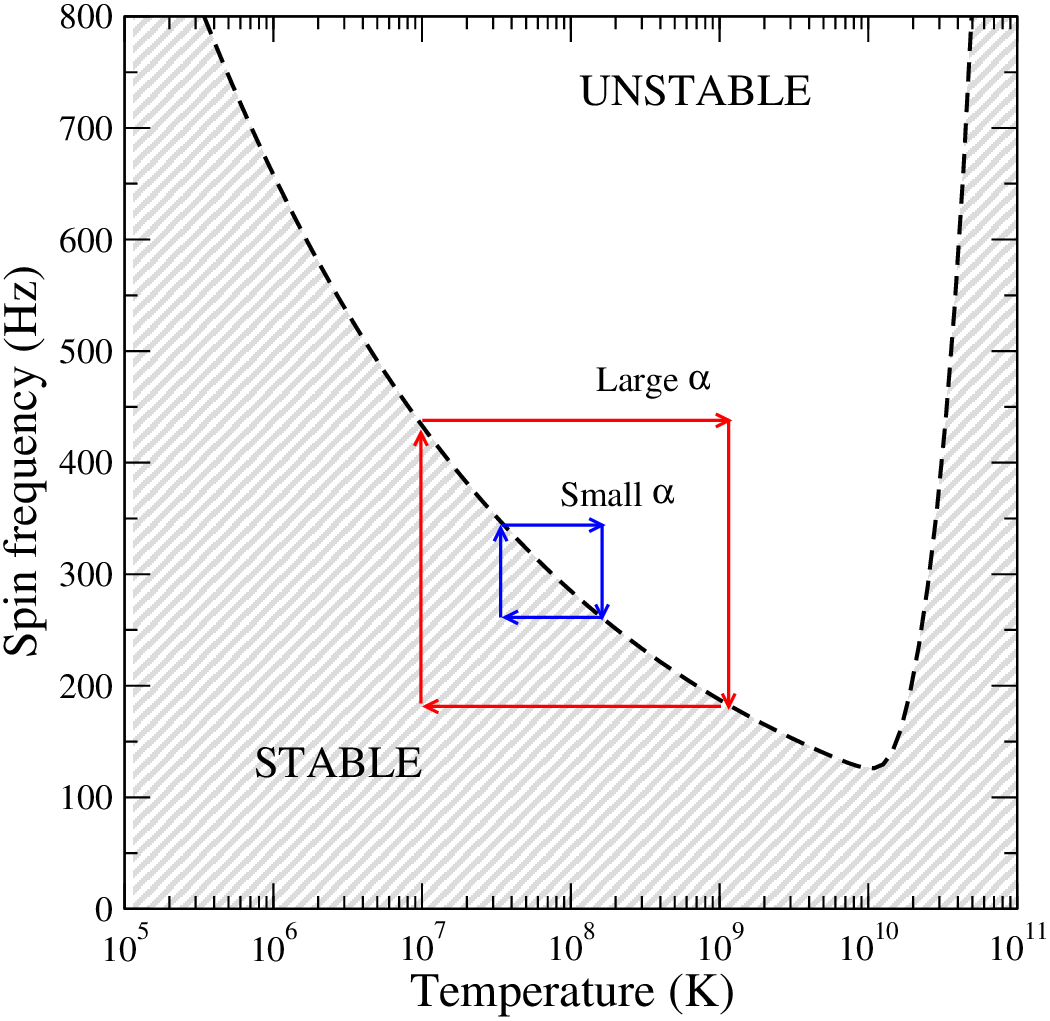}}
%
%
\caption{The r-mode instability window for a 1.4 $M_\odot$, $R=10$ km NS. The equation of state is an $n=1$ polytrope and we consider the `minimal' model described in the text. In the shaded region the mode is stable, and unstable above it. At low temperatures the main source of damping is  given by Ekman pumping the crust-core boundary, while at high temperature bulk viscosity gives the main contribution. We also show,  both for small and large amplitudes $\alpha$ of the mode, the cycle that an accreting system would follow in the temperature-frequency plane.}
\label{window1}
\end{figure}

In order to understand how a system evolves in the presence of an unstable r-mode it is useful to consider an 'instability' window, i.e. a region in parameter space in which the mode is unstable. In order to do this we will fix the mass of the star at $M=1.4 M_\odot$ and the radius at $R=10$ km and examine the region in the spin frequency vs temperature plane in which the mode can go unstable. The boundary of this region corresponds to the points in which the damping and driving timescales are equal, i.e. to the solutions of
\be
\frac{1}{\tau_{gw}}=\sum_i \frac{1}{\tau_{V i}}
\ee
where $\tau_{V i}$ is the viscous damping timescale for process $i$ acting in the star. The result is shown in figure \ref{window1}. 

Phenomenological evolution equations for the frequency $\Omega$ of the star, the mode amplitude $\alpha$ and the thermal energy $E_T$ of the star take the form \cite{Owen98}:
\beq
\frac{d \Omega}{d t}&=&-\frac{2\Omega}{\tau_V}\frac{\alpha^2 Q}{1+\alpha^2 Q}\\
\frac{d \alpha}{dt}&=& -\frac{\alpha}{\tau_{gw}}-\frac{\alpha}{\tau_V}\frac{1-\alpha^2 Q}{1+\alpha^2 Q}\\
\frac{d E_T}{dt}&=&-\dot{E}_\nu+\dot{E}_V
\eeq
where $\tau_V$ is the damping timescale due to the dominant viscous mechanisms, the dimensionless parameter $Q=9.4\times 10^{-2}$ for an $n=1$ polytrope \cite{Owen98} and $\dot{E}_\nu$ is the luminosity due to the modified URCA process \cite{Shapiro}:
\be
\dot{E}_\nu=1.1\times 10^{32} \left(\frac{M}{1.4 M_\odot}\right)\left(\frac{\rho}{10^{15}\mbox{g/cm$^3$}}\right)^5\left(\frac{T}{10^8 \mbox{K}}\right)^2 erg/s
\ee
with $\rho$ the density, while $\dot{E}_V$ is the reheating due to shear viscosity as the mode grows\cite{NilsRev}:
\be
\frac{dE}{dt}=3.3\times 10^{-2}\frac{\alpha^2\Omega^2MR^2}{\tau_{sv}}\label{heat}
\ee

A typical accreting NS, with a core temperature of around $10^8$ K will be spun up by accretion into the unstable region. The r-mode then rapidly grows to large amplitude, resulting in fast heating. Eventually the thermal runaway is halted by cooling due to neutrino emission and the star simply spins down due to GW emission, re-entering the stable region.  It will then cool and eventually start the cycle again, as shown in figure \ref{window1}. 

However the above equations are only valid as the mode is growing, and in reality the amplitude will saturate at a value $\alpha_s<<1$ due to non-linear couplings to other modes. We can see from equation \ref{heat} that this saturation amplitude plays a critical role in determining the amount of heating and ultimately how far into the instability window a NS can move.

If the saturation amplitude is very large ($\alpha_s\approx 1$) the system will move well into the unstable region, but the evolution will be fast and the duty cycle very short \cite{Levin, Spruit}, less than $\approx 1\%$ . Conversely, if the mode saturates at low amplitude ($\alpha\approx 10^{-5}$), as suggested by calculations of non-linear couplings to other modes \cite{IRA1}, the duty cycle is much longer but the system will remain close to the instability curve \cite{Heyl}. In both scenarios it is highly unlikely to observe a system far inside the unstable region.

\section{Observational constraints on the instability window}
\label{observe}
\begin{figure}[th]
\centerline{\includegraphics[scale=.42]{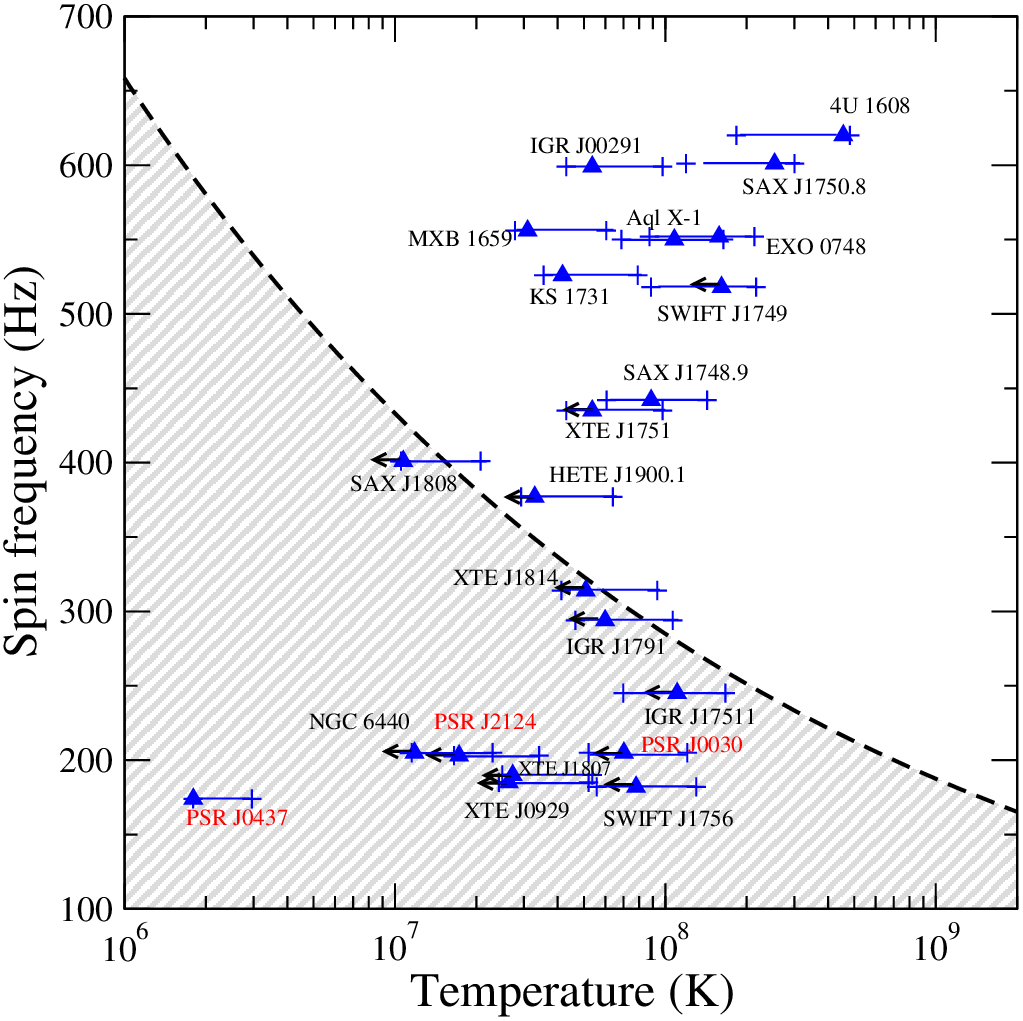}}
%
%
\caption{Comparison between the observed spins and core temperatures of NS and the r-mode instability window in our `minimal' model, for a $M=1.4 M_\odot$ and $R=10$ km star, described by an $n=1$ polytrope. The error bars are due to the uncertainty on the composition of the outer layers of the star. There is clearly a large number of sources in the unstable region, while our expectation is that they would rapidly be spun out for large saturation amplitudes of the mode, or remain close to the instability curve if $\alpha_s$ is small.}
\label{sources}
\end{figure}

The main conclusion of the previous section is that one would not expect to see systems in the instability window, and that they should either be close to the instability curve or well below it. We can test this prediction for our `minimal' model by populating the window with observations of spins and temperature in LMXBs. The spins of NSs in LMXBs can either be measured directly in systems for which coherent X-ray pulsations are detected, or inferred from the frequency of oscillations seen in thermonuclear type I X-ray bursts \cite{Ale10, AleAnna}. Temperatures, on the other hand, require some degree of modelling, given that we are interested in the {\it core} temperature. Surface temperatures of NSs can be estimated from black body fits to spectra of LMXBs in quiescence. To obtain the core temperature one can then assume that the interior is roughly isothermal, and model the exterior layers of the NS to obtain a relation between the surface temperature and the temperature at the base of the envelope\cite{Pot}. 

In figure \ref{sources} we see the result of populating our `minimal' instability window for a $M=1.4 M_\odot$ and $R=10$ km NS, with data from LMXBs, as obtained by Haskell, Degenaar and Ho \cite{HDH}. The error bars in the figure are due to the uncertainty in the composition of the outer layers. It is quite obvious that there is a large number of sources inside the instability window, in a region that would not be permitted. Given the limited sample of systems for which one has measurements of both spins and temperatures (22 in the case of figure \ref{sources}) one would not expect to find any in the unstable region \cite{Heyl}. The conclusion is quite robust and was found to hold even for different masses and equations of state \cite{Siminrmode}, and to be consistent with more detailed modelling of the core temperatures of LMXBs \cite{Wynnlett}.

This problem clearly shows that the `minimal' model we described, and that is often considered in NS physics, needs to be re-evaluated and additional physics must be included. In the following we describe some of the main mechanisms that could make our theoretical understanding consistent with observations.

\section{Additional damping mechanisms}

First of all we examine the possibility that additional physics, such as exotic particles in the core or strong superfluid mutual friction, may provide strong damping at low temperatures, and modify the shape of the instability window. Let us thus outline some of the most promising mechanisms.

\subsection{Mutual friction}
\label{MF}

Neutrons in the interiors of NS are thought to pair and form a large scale superfluid condensate that can oscillate independently from the proton-electron fluid, to which it is only very weakly coupled \cite{Baym}. R-modes in superfluid neutron stars have been studied in detail by several authors \cite{LM00,LY03,BH09, Andrea09}, who have found that two families of modes can now exist, one in which the fluids are mainly co-moving and another in which neutrons and protons are counter moving which, however, only exists in non-stratified stars \cite{BH09}. Rotation `mixes' the modes and, to second order in rotation, the standard co-moving r-mode also has a counter-moving component which, for millisecond spin-periods such as those we are considering, can be quite large and lead to damping via vortex mediated superfluid mutual friction. The strength of the mutual friction is usually quantified by a dimensionless parameter $\mathcal{R}$ which encodes the microphysics that gives rise to the effect. The coupling timescale between the superfluid neutrons and the electrons then scales as $\tau\approx 1/2\Omega\mathcal{R}$ for $\mathcal{R}<<1$ as is generally the case in NSs. 

\begin{figure}[th]

{\includegraphics[scale=.35]{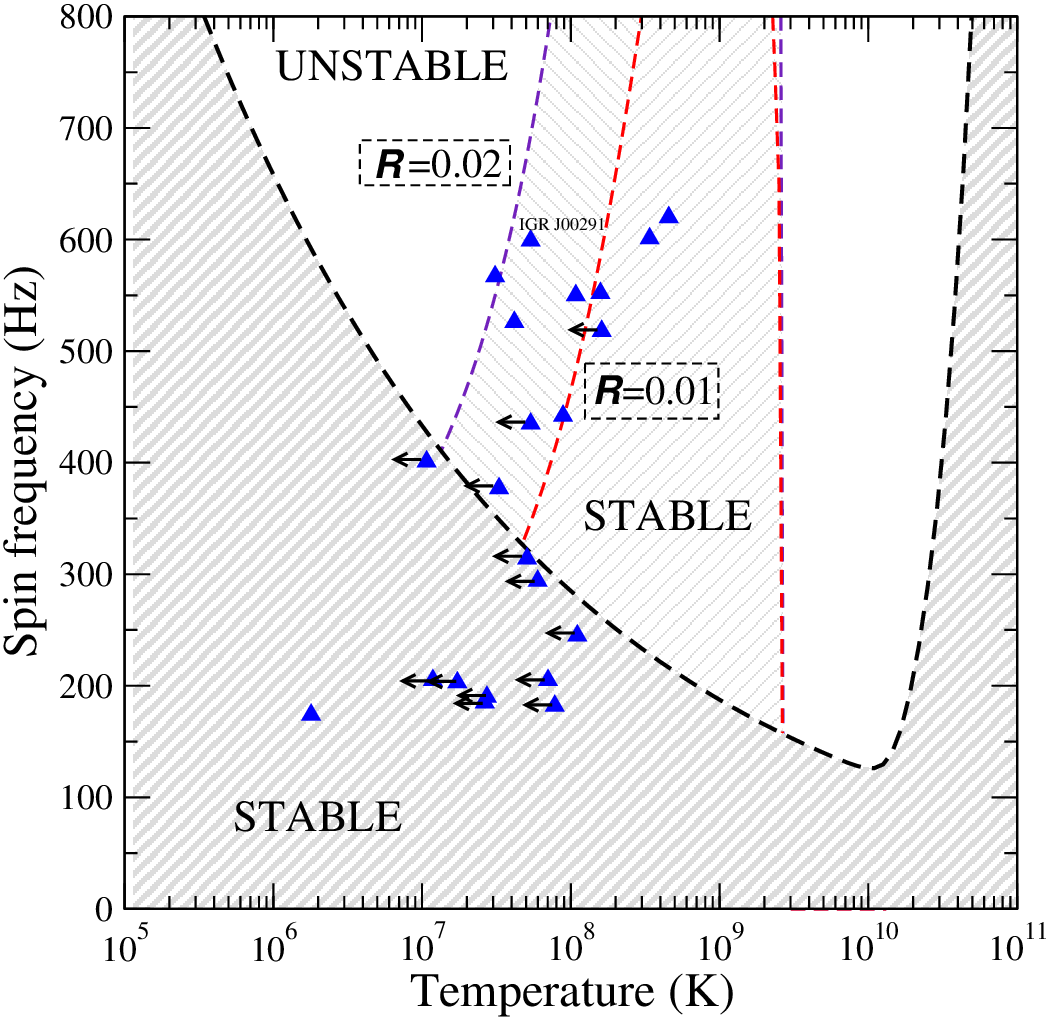}\includegraphics[scale=.35]{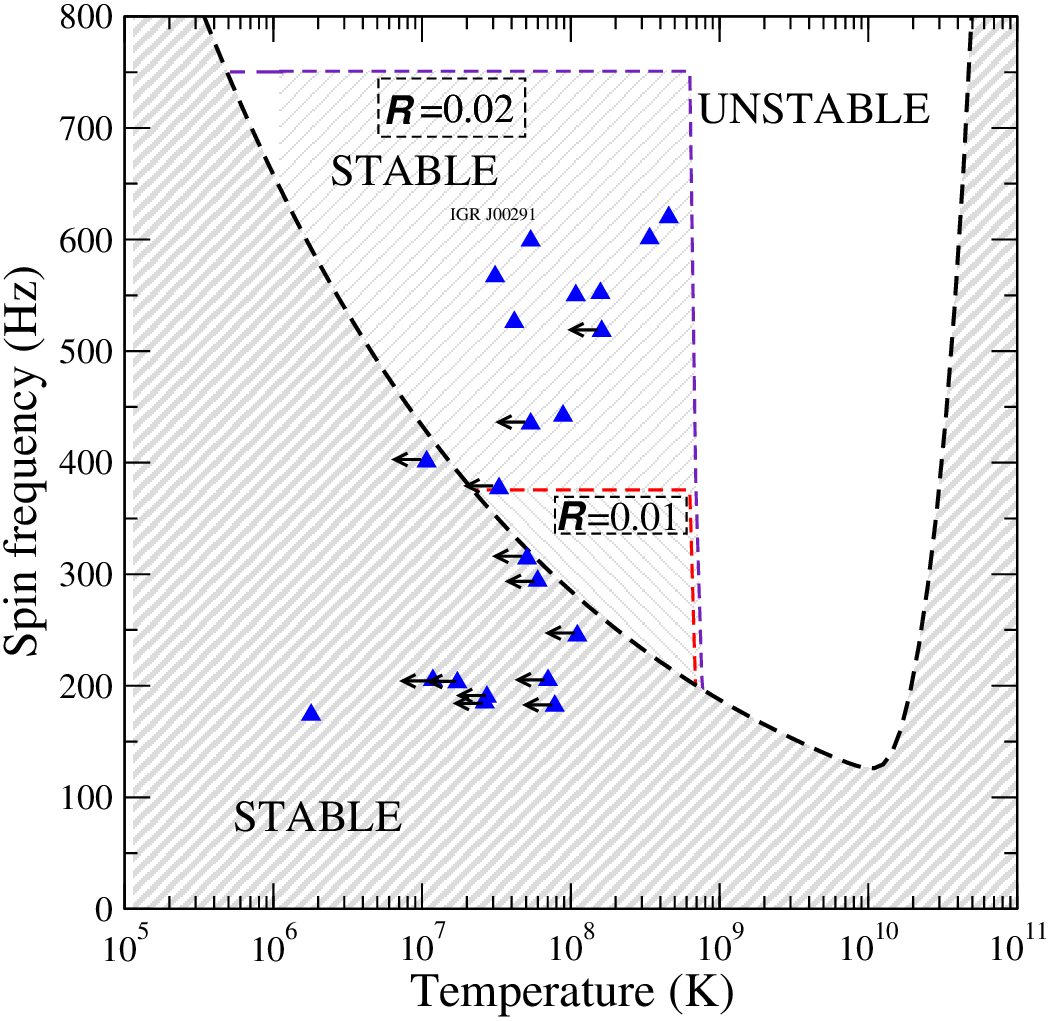}}

\caption{The r-mode instability window for strong mutual friction. There is a considerable difference in the windows depending on the superfluid pairing gaps that are used: we show the 'strong' (left) and 'weak' (right) models described in Haskell et al. 2009\cite{BH09}. In both cases the theoretical model agrees with observations for $\mathcal{R}\approx 0.01$, which is in the possible range for the vortex flux tube cutting mechanism.}
\label{windowMF}
\end{figure}
The most commonly considered mutual friction mechanism is the scattering of electrons of magnetised vortex cores \cite{AlparMF, Sidery06}, which gives $\mathcal{R}\approx 10^{-4}$. In this case the stabilising effect on the r-mode is weak and the instability window is not altered \cite{BH09, passamontirmode}.  It is thus interesting to ask how large the parameter $\mathcal{R}$ needs to be for the instability window to be consistent with observations. As can be seen in figure \ref{windowMF} the instability curve depends quite strongly on the superfluid pairing gaps that are used (which can also have an impact on the strength of shear viscosity \cite{Tolos1}), but is still generally consistent with observations for $\mathcal{R}\approx 10^{-2}$. This value could be consistent with what is expected if the protons in the outer core form a type II superconductor and the superfluid vortices `cut' through superconducting flux tubes \cite{Link03}, in which case one expects \cite{saturation}:
\be
\mathcal{R}\approx 2.5\times 10^{-3} \left(\frac{B}{10^{12}\mbox{ G}}\right)^{1/4}
\label{rrb}
\ee
where $B$ is the strength of the magnetic field. The exterior dipole component of the magnetic field in LMXBs is generally inferred to be in the region $B\approx 10^8-10^9$ G, so the interior field would have to be significantly stronger for this mechanism to work. It is important to note though that in equation (\ref{rrb}) the mutual friction coefficient $\mathcal{R}$ is velocity dependent, and we shall see in the following that this can have important consequences for the saturation amplitude of the r-mode.

\subsection{Exotica in the core}

\begin{figure}[th]

\centerline{\includegraphics[scale=.42]{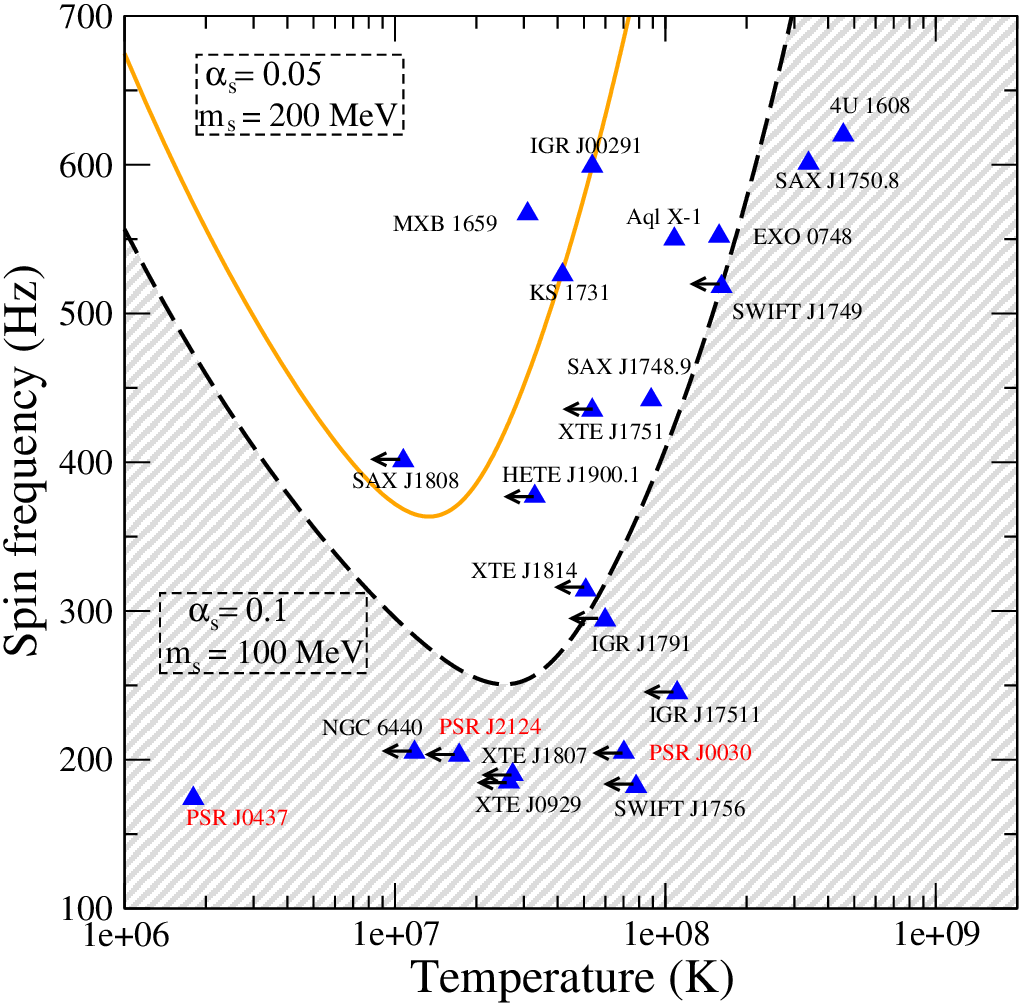}}

\caption{The r-mode instability window in the case of a strange quark star with $M=1.4 M_\odot$ and $R=10$ km \cite{HDH}. The exact shape of the window depends slightly on the choice of model parameters, such as the value of the strong coupling constant $\alpha_s$ and the mass of the strange quark $m_s$ but the qualitative features, namely the increase of viscosity around $T=10^8$ K are fairly insensitive to the exact microphysical description of the quark core \cite{Alford12}.}
\label{quark}
\end{figure}

If exotic particles, such as hyperons or deconfined quarks, are produced in the core of the NS this generally leads to an increase of the bulk viscosity at low temperatures which significantly alters the instability window. In figure \ref{quark} we show the instability window in the case of a strange star with $M=1.4 M_\odot$ and $R=10$ km \cite{HDH}. The main point to note is that the increase in bulk viscosity at around $T\approx 10^8$ K can explain the presence of several systems in this region. The window depends slightly on the model that has been chosen for the shear and bulk viscosity, and on the parameters of the theory, however Alford and collaborators \cite{Alford12} have shown that the main features of the instability window are remarkably insensitive to the exact microphysical description of matter, and models with a sizeable ungapped quark are generally consistent with observations \cite{Alford14, Alfordlett}. Note, however, that if quarks in the core are paired in the CFL phase, viscosity is much lower and the instability window is essentially that of the `minimal' model \cite{cflletter}. The appearance of hyperons in the core can also lead to an increase in bulk viscosity at low temperatures \cite{Haskellhyperon, Nayyar}.

If r-modes in LMXBs are stabilised by strong bulk viscosity at low temperatures, this poses, however, a new theoretical challenge, as one would expect the systems to cool and spin down along the instability curve, after accretion ceases. This is at odds with the observations of several fast (with spin frequencies up to $700$ Hz) millisecond radio pulsars. This issue is not present if shear viscosity is enhanced at high temperatures, which may be the case in the presence of pion condensation \cite{pions}.

\subsection{The crust/core interface}

In our `minimal' model we have included the effect of damping due to a viscous boundary layer at the crust/core interface. We have accounted for the possibility that the crust may not be completely rigid but participate in the oscillation by introducing a slippage parameter \cite{Ek1, Ek2} $\mathcal{S}$ that is essentially the ratio between the jump in velocity at the crust core interface and the mode velocity. For small values of $\mathcal{S}$ the crust can oscillate freely, while for $\mathcal{S}=1$ the crust is completely rigid, and in this case all the observed systems would be r-mode stable. In our model we have set $\mathcal{S}=0.05$, however the crust does not respond in the same way at all frequencies, as resonances between the r-mode and torsional oscillations of the crust are possible. This leads to the possibility that the rigidity of the crust, i.e. the parameter $\mathcal{S}$, may be frequency dependent \cite{Levincrust, Wynnlett}, and more detailed modelling of the coupling between crust and core is necessary to understand if this mechanism is consistent with observations of temperatures and spins in LMXBs.

Note, however, that if the crust/core transition is not sharp, but more gradual and proceeds via several phase transitions that give rise to the so-called pasta phases, then viscosity would be much weaker and by mainly due to standard shear viscosity \cite{Will1}.

\subsection{Interactions with superfluid modes}

Another interesting possibility that was introduced by Gusakov and collaborators \cite{GusakovLett} is that of interactions between `superfluid' inertial modes and the standard r-mode, which can take place at fixed `resonance' temperatures and lead to `spikes' in the instability window as the mode changes character \cite{LY03, Guspaper}. This is due to the fact that the `superfluid' inertial modes are counter-moving oscillations of the fluid already to leading order in the rotation rate of the star $\Omega$. For these `superfluid' modes one has for the counter-moving component of the velocity $\mathbf{w}_s\approx O(\Omega)$, compared to $\mathbf{w}_n\approx O(\Omega)^3$ for the `normal' r-mode.
As counter-moving motion is the main driver behind mutual friction, a large $\mathbf{w}$ at leading order in rotation leads to a short mutual friction damping timescale, such that for a `supefluid' mode one has $\tau_{MF}^s\approx O(\Omega)$, rather than $\tau_{MF}^n\approx O(\Omega)^5$ for a normal r-mode.
Mutual friction is the main viscous process acting on  `superfluid' inertial modes and provides efficient damping, while the coupling to GW emission is much weaker \cite{fmode1}. The r-mode is thus stabilised by the interaction with these modes.

This scenario makes an observational prediction, namely that one should uncover a population of HOt and Fast Non-Accreting Rotators (HOFNARs) that are essentially `hot windows', i.e. systems in which the NSs has been re-heated by the r-mode instability and appears as a thermally emitting isolated system after the accretion phase is over \cite{HOFNAR}. Future X-ray observations could thus confirm the viability of this scenario.

\section{Saturation amplitudes}

Up to now we have described mechanisms that can enhance the viscosity in the temperature and frequency range of LMXB observations, and modify the instability window so that these systems are r-mode stable. The `minimal' model for this instability window could, however, be made consistent with observations if the saturation amplitude of the r-mode is so small that the instability is indeed present in all the observed systems, but at such a low-level as to not impact on the spin or thermal evolution of the stars. 

This scenario requires very small amplitudes $\alpha_s\approx 10^{-9}-10^{-6}$ \cite{HDH, Siminrmode}, generally much smaller than the values obtained for saturation due to non-linear couplings with other inertial modes \cite{IRA1}, which lead to saturation amplitudes $\alpha_s\approx 10^{-5}$. It has been suggested \cite{IRAsmall} that if the crust/core transition is smeared out, for example due to the appearance of pasta phases, as previously discussed, then viscosity will be much weaker and allow for inertial modes to grow and saturate the r-mode at amplitudes $\alpha_s\lesssim 10^{-6}$. These calculations ignore, however, the effect of mutual friction which can quite efficiently halt the growth of inertial modes coupled to the r-mode.

If the core of the star is in a type II superconducting state, however, mutual friction due to superfluid vortices cutting though superconducting flux tube may effectively saturate the r-mode at low amplitudes \cite{saturation}. As we have already discussed if vortices can cut through flux tubes this leads to strong mutual friction, with $\mathcal{R}\approx 2.5\times 10^{-3} B_{12}^{1/4}$ \cite{saturation}, this, however, is not always the case, as the large energy penalty of cutting leads to vortices `pinning' to flux tubes for counter-moving velocities $\mathbf{w}_p$ less than
\be
|\mathbf{w}|_p\approx 1.5 \times 10^{4} \left(\frac{B}{10^{12} \mbox{G}}\right)^{1/2} \;\;\mbox{cm/s}
\ee
where $B$ is the macroscopic core magnetic field. If vortices are pinned and cannot cut through flux tubes there is essentially no mutual friction damping and the r-mode can grow (provided other sources of viscosity are not suppressing the instability). As the mode grows, however, the amplitude of the counter-moving component of the velocity $\mathbf{w}$ also grows, according to 
\be
|\mathbf{w}|\approx \lambda_0 \alpha \left(\frac{r}{R}\right)^2 \left(\frac{\Omega}{\Omega_K}\right)^2 R\Omega
\ee
with $\Omega_K$ the Keplerian breakup frequency of the star and $\lambda_0$ a spin independent parameter that lies in the range $\lambda_0\approx 0.1-1$\cite{BH09, saturation}.
When $\alpha$ grows large enough it is thus possible to enter the regime in which $w>w_p$ and vortices are forces through flux tubes. At this point strong mutual friction due to cutting will take over and damp the mode, effectively halting the growth of the instability at an amplitude
\be
\alpha_{pin}\approx 10^{-6} \left(\frac{\lambda_0}{0.1}\right)^{-1}\left(\frac{\nu}{500\mbox{Hz}}\right)^{-3}\left(\frac{B}{10^{8} \mbox{G}}\right)^{1/2}
\ee
with $\nu$ the spin frequency of the star in Hz. In many cases this amplitude can be smaller than the limit set by non-linear mode couplings \cite{saturation}. Even lower saturation amplitudes of $\alpha\approx 10^{-10}$ are possible in hybrid stars due to periodic phase conversion at the interface with the exotic core \cite{Alfordsaturation}.

Another possibility is that the r-mode oscillation will wind up the interior magnetic field of the star to the produce a strong toroidal component, which could then rapidly suppress the instability if the internal magnetic field of the star is of order $B\approx 10^{10}$ G \cite{lucy1, lucy2, lucy3, drago}. Note that this mechanisms is expected to be active only if the r-mode is driven unstable by GW emission \cite{Diff1, Diff2}, and if the amplitude is not growing the effect of the magnetic field will be negligible for the field strengths expected in LMXBs \cite{Abba12, Chirenti13, Asai15}.

The small saturation amplitudes predicted in this section would all lead to GW emission well below the detectability level for current and next generation GW detectors, for which only emission at the level required to balance the accretion torque and explain the observed spin periods would be possible, although challenging, to detect \cite{Watts08}. Emission at that level would, however, be inconsistent with the observed temperatures of LMXBs, as it would heat the stars up more than is observed \cite{Wynnlett, HDH}. It is thus likely that GWs from r-modes in accreting neutron stars in LMXBs will be very challenging to detect, with young NSs being a much more promising target \cite{Alfordyoung, Alfordlett}.

\section{Electromagnetic observations of r-modes}

Recently Strohmayer \& Mahmoodifar \cite{Simin1,Simin2} have conducted targeted searches for oscillations in the X-ray light curves of several LMXBs and discovered candidate features in two of them, XTE J1751-305 and 4U 1636-536. In table \ref{modestab1} we show the oscillation frequencies and characteristics of the two systems.

For the first, XTE J1751-305 the oscillation frequency has been interpreted as the {\it rotating} frame frequency of the mode, as it is thought that we are observing modulations of the X-ray emitting hotspots due to a mode on the surface of the star \cite{numata}. The best candidate for this mechanism is a surface g-mode \cite{BC98, PB04}, as it can produce large modulations. However r-modes are also likely candidates, especially due to the fact that in a stratified star there are interactions (avoided crossings) between the g-modes and inertial modes such as the r-modes, that can modify the eigenfunctions and the nature of the oscillations \cite{passastrati}.
If the oscillation is interpreted as a global r-mode the amplitude required to explain the observed modulation of the light curve is $\alpha\approx 10^{-3}$ which is very large, and would lead to the system spinning down due to GW emission, which is not observed \cite{NilsObserve}. This problem is somewhat alleviated if one accounts for a rigid crust, in which case interactions with crustal oscillations can amplify the amplitude at the surface by up to two orders of magnitude \cite{YL01, unnonew}, leading to amplitudes in the bulk of the star of $\alpha\approx 10^{-5}$. This amplitude would still be too large to be consistent with the observed temperature of the star \cite{Siminrmode}, which would require $\alpha\approx 10^{-8}$, but suggests that further detailed modelling is necessary.

\begin{table}[pt]
\tbl{The spin frequency of the star ($\nu$) and observed mode frequency for the two systems in which an oscillations was detected by Strohmayer \& Mahmoodifar \cite{Simin1, Simin2}. We also show the ration $\kappa$ between the mode frequency and the spin frequency.\label{modestab1}}
{\begin{tabular}{@{}cccc@{}} \toprule
System & $\nu$ (Hz) & Mode  frequency (Hz) & Ratio $\kappa$  \\ \colrule
XTE J1751-305 & 435&  249.33 & 0.5727597\\
4U 1636-536 & 582& 835.6440& 1.43546 \\
 \botrule

\end{tabular}}

\end{table}

In the second system, 4U 1636-536, the oscillation appears during a super burst, which is thought to be a thermonuclear burst due to unstable burning of a carbon layer formed by the ashes of a regular type-I X-ray burst. In this case, consistently with the idea that the burning layer spreads rapidly and the emission is coming from the whole surface, the observed frequency is interpreted as the {\it inertial} frame frequency of the mode. As for the previous system, several modes could oscillate at the observed frequency, with g-modes and r-modes being the main candidates. 4U 1636-536, however, is not observed as an X-ray pulsar, and there is no measurement for its long term spin-down rate. One cannot thus rule out the possibility of a large amplitude r-mode spinning down the star.

Finally note that both these events took place before the current LIGO science runs (in 2002 and 2001 respectively). If such an event were to repeat itself while advanced detectors are taking data a large amplitude r-mode would be detected\cite{NilsObserve}. Needless to say a simultaneous X-ray and GW observation of a NS oscillation would allow us an unprecedented insight into the interior dynamics of these systems and inaugurate the era of NS asteroseismology.

\section{Conclusions}

In this article I have reviewed the main theoretical aspects of the GW driven r-mode instability in rapidly rotating neutron stars and described what the instability window (i.e. the region in the frequency vs. core temperature plane in which the r-mode is unstable) is predicted to be in a `minimal' NS model that assumes a core of neutrons, protons and electrons, with no exotica (such as deconfined quarks), dynamically significant superfluid degrees of freedom or magnetic fields. 

The predictions of such a model can be compared to observations of spins and temperatures in LMXBs. The expectation is that no system should fall inside the unstable region. If the saturation amplitude of the mode is small ($\alpha\approx 10^{-5}$) as predicted by calculations of nonlinear couplings to other modes \cite{IRA1}, then the NS will never depart significantly from the instability curve. On the other hand, if the saturation amplitude  is large ($\alpha\approx 1$) the system can enter well into the instability region, but the timescale to exit it will be very short, leading to a very low duty cycle and probability of observing a system in this phase.

In section \ref{observe} I show that the predictions of the `minimal' model are not consistent with observations \cite{HDH}, as many observed systems would fall in the unstable region. Despite uncertainties on the mass of the NS and on the composition of the outer layers, this qualitative conclusion is robust and indicates the need to include additional physics in our model. There are essentially two theoretical solutions to this observational puzzle: either there are additional physical mechanisms giving rise to strong viscosity at low temperature, and making the observed systems stable, or the systems are indeed unstable, but the r-mode saturates at such a low amplitude that it does not impact on the spin or thermal evolution of the system.

Additional viscosity could be due to hyperons \cite{Nayyar, Haskellhyperon} or deconfined quarks in the core \cite{Alford12, HDH}, strong mutual friction due to vortex/flux tube cutting \cite{BH09}, resonances with crustal torsional modes \cite{Levincrust, Wynnlett} or to interactions between the r-mode and superfluid inertial modes \cite{GusakovLett}. Very small saturation amplitudes, on the other hand, could be possible if the growth of the mode is halted before non-linear couplings set in, by processes such as vortex/flux tube cutting \cite{saturation}, periodic phase transitions in hybrid stars \cite{Alfordsaturation} or by winding up a strong toroidal component of the magnetic field \cite{lucy1}.

Finally I have reviewed recent observations of two oscillation modes detected in the X-ray light curve of the LMXBs  XTE J1751-305 and 4U 1636-536 \cite{Simin1, Simin2}. In both cases several modes could lead to the observed frequencies, with the most likely being surface g-modes or r-modes. For the first system the observed frequency is likely to correspond to the rotating frame frequency of the mode, as the surface oscillation perturbs the emitting hotspot. For the second system, however, the oscillation is detected during a thermonuclear super burst, in which burning is likely to be occurring over the whole surface of the star. In this case the observed frequency would correspond to the inertial frame frequency of the mode.

Both modes can be interpreted as r-modes, however to explain the observed modulation of the X-ray fluxes the corresponding amplitudes would have to be large ($\alpha\approx 10^{-3}$). Although the surface amplitude can be amplified by up to a factor of 100 by interactions with crustal modes \cite{unnonew}, the amplitude is still unrealistically large for XTE J1751-305, as the star would have to be hotter than observed, and spin-down faster than observed \cite{NilsObserve}. In the case of 4U 1636-536, however, a large amplitude r-mode cannot be excluded.

If such an event were to repeat itself while advanced GW detectors are operating, a large amplitude r-mode could be detected, allowing for the tantalising possibility of a coincident electromagnetic and GW detection, and truly opening the era of NS asteroseismology.

\section*{Acknowledgements}

I acknowledge support from the Australian Research Council (ARC) via a DECRA fellowship. 

\bibliographystyle{ws-ijmpe}
\bibliography{rmodes}







\end{document}